\documentclass[camera]{gji2}

\usepackage{times} 
\usepackage[latin1]{inputenc} 
\usepackage[english]{babel}
\usepackage{graphicx} 
\usepackage{amssymb}
\usepackage{natbib}

\newcommand{\mygraphicspath}{./}

\newcommand{\ttc}{T$_2$C}

\usepackage{changebar}

\newcommand{\mb}[1]{\mbox{#1}}

\newcommand{\rc}{\ensuremath{{(\rho c)_c}}} 
\newcommand{\Ta}{\ensuremath{{T_a}}} 

\newcommand{\inlm}[4]{\int\limits_{#1}^{#2} #3 \; #4}

\newcommand{\uipl}{\ensuremath{\mb{ %
        W}/\mb{  m}}}

\newcommand{\uic}{~\ensuremath{\mb{W}/(\mbox{m K})}}
\newcommand{\ucb}{\ensuremath{\left[\frac{\mb{ %
          W}}{\mbox{m K}}\right]}}

\newcommand{\uid}{\ensuremath{\cdot10^{-7}\mb{ %
        m}^2/\mb{s}}}
\newcommand{\udb}{\ensuremath{\left[10^{-7}\frac{\mb{ %
          m}^2}{\mb{  s}}\right]}}

\newcommand{\uiql}{\ensuremath{\mb{ %
        J}/\mb{  m}}}
\newcommand{\uqlb}{\ensuremath{\left[\frac{\mb{ %
          J}}{\mb{  m}}\right]}}

\newcommand{\uirc}{\ensuremath{\cdot10^6\mb{ %
        J}/(\mb{K m}^3)}}
\newcommand{\urcb}{\ensuremath{\left[10^6\frac{\mb{ %
          J}}{\mb{  K m}^3}\right]}}

\begin{document}

\title[Inversion of heat flow measurements]{Inversion of marine heat
  flow measurements by expansion of the temperature decay function}

\author[A. Hartmann and H. Villinger]{A. Hartmann \thanks{Present address:
    RWTH Aachen, Angewandte Geophysik, Lochnerstrasse 4--20,
    52056 Aachen, Germany, email:~a.hartmann@geophysik.rwth-aachen.de} and H.
  Villinger  \\ Fachbereich Geowissenschaften, Universit\"{a}t Bremen \\
  Postfach 330440, D-28213 Bremen, Germany \\ email:
  vill@uni-bremen.de}

\maketitle

\begin{summary}
  Marine heat flow data, obtained with a Lister-type probe, consists
  of two temperature decay curves, frictional and heat pulse decay.
  Both follow the same physical model of a cooling cylinder. The
  mathematical model describing the decays is nonlinear as to the
  thermal sediment parameters thus a direct inversion is not possible.
  To overcome this difficulty, the model equations are expanded using
  a first order Taylor series. The linearised model equations are used
  in an iterative scheme to invert the temperature decay for
  undisturbed temperature and thermal conductivity of the sediment.
  The inversion scheme is tested first for its theoretical limitations
  using synthetic data. Inversion of heat flow measurements obtained
  during a cruise of R/V SONNE in 1996 and needle probe measurements
  in material of known thermal conductivity show that the algorithm is
  robust and gives reliable results. The program can be obtained
  from the authors.
\end{summary}


\section*{Introduction}
\label{sec:intro}

The growing interest in global energy and geochemical fluxes in the
oceans and the availability of multi-penetration heat flow probes have
has led to increased attention in marine geothermal studies, which are
mainly focused on the sediment covered areas of ridges and ridge
flanks. Studies of accretionary wedges that incorporate heat flow
investigations also reflect the need for additional constraints to
model and understand the dewatering process of accreted sediments.
This increased interest in marine heat flow data has helped to improve
the existing measurement technique in two ways: The violin bow type
heat probe instrument, as described in \citet{Hyndman79}, has evolved
over two decades of intensive use to a mature, mechanically robust
instrument which now can be used in a routine way. Rapid
electronic development led to an increased temperature resolution of
1~mK and allowed a larger number of sensors to be mounted on one
string due to increased storage capacity.  Both developments now permit 
multi-penetration deployments (measurements in a pogo-style
fashion) of up to 24 hours per station.  Advances in marine navigation
now permit very detailed studies of regional processes.  Worldwide
coverage of the Global Positioning System (GPS) and Differential GPS
help enormously in station keeping during a measurement.  High
accuracy in positioning of the probe is achieved by using Long- or
Short-Baseline navigation \citep{Jones1999}.

\begin{figure}
  \begin{center}
    \includegraphics[width=5cm]{\mygraphicspath 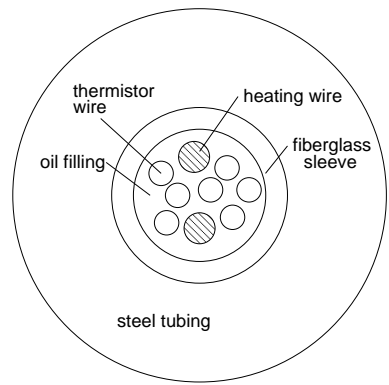}
     \caption[Cross section of the sensor tube.]{Cross section of the
       sensor tube, housing the temperature sensors (thermistors) and
       the heating wires for in situ thermal conductivity measurements
       (see also Table \ref{tab:therm_parm_sn_tube}).}
    \label{fig:cross_section_sn_tube}
  \end{center}
\end{figure}

\begin{table}
  \begin{center}
    \begin{tabular}{p{1.5cm}|c|c|c|p{1.9cm}} \hline
      Part & Layer & $k$ & \rc\ &  Material \\
        & & \ucb & \urcb & \\ \hline \hline
      Sensor string & 1  & 0.2 & 2.14 & copper, oil, glass \\ \hline
      Heating wire & 2  & 13.2 & 3.50 & Ni-Chrome \\ \hline
      Oil filling & 3  & 0.2 & 1.60 & oil, plastic \\ \hline
      steel tube & 4  & 37.5 & 3.9 & steel \\ \hline
      \end{tabular}
      \caption[Thermal parameters of the sensor tube]{Thermal
        parameters of the sensor tube of
        figure~\protect\ref{fig:cross_section_sn_tube}
        (after~\protect\citet{Nagihara93}).}
    \label{tab:therm_parm_sn_tube}
  \end{center}
\end{table}

Figure~\ref{fig:cross_section_sn_tube} shows a simplified cross
section of a sensor string used by the Heat Flow Group of the
University of Bremen (Germany) and the Pacific Geoscience Centre
(Sidney, B.C., Canada). The sensors for temperature measurements in
the sediment are housed in a hydraulic steel tube with an outer
diameter of 8~mm and an internal diameter of 5~mm.  Thermal parameters
are shown in Table~\ref{tab:therm_parm_sn_tube} (after
\citet{Nagihara93}). The sensor string itself contains heating wires
for in situ thermal conductivity measurements and wires leading to the
thermistors for temperature measurements. The tube is oil-filled to
improve the thermal contact between the temperature sensors and the
tube.

\begin{figure}
  \begin{center}
    \includegraphics{\mygraphicspath 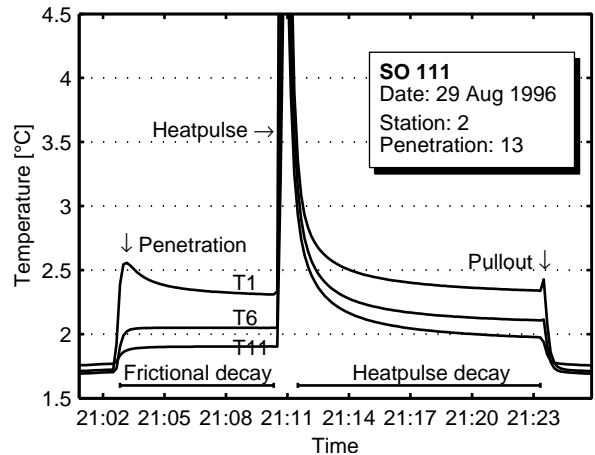}
    \caption[Example of a heat flow measurement.]{Example of a heat
      flow measurement. Only three out of 11~sensors are shown.}
    \label{fig:penexample}
  \end{center}
\end{figure}

In order to illustrate the steps involved to
process a heat flow measurement, Fig.~\ref{fig:penexample} shows a
typical data set.
Measurements of undisturbed sediment temperature and conductivity
follow the pulsed needle probe method \citep{Lister70}.  When the
sensor string penetrates the sediment the friction between sensor tube
and sediment creates heat resulting in a temperature rise.  The
following temperature decay is recorded at 10~s sample rate for a
preset time span (7~minutes), after which a calibrated heat pulse of
20~s length is fired.  The heat pulse decay is monitored for at
least 7~minutes until the probe is pulled out of the sediment and the
ship moves to the next measurement position with the probe about 
200--300~m above the seafloor (pogo-style measurements).

The processing of the raw measurements requires three steps: (1)
determine undisturbed sediment temperatures from frictional decay (2)
correct heat pulse decay for the remaining effect of the frictional
decay and (3) calculate thermal conductivities from heat pulse decay.
The basic design of the processing of heat flow measurements is
outlined in \citet{Hyndman79} which was basically a manual procedure
based on the work of \citet{Lister70} and \citet{Lister79}.

The increasing number of measurements per cruise in the past decade
required a processing scheme which could easily be implemented on a
personal computer for automated processing of a large number of
measurements.  \citet{Villinger87} published a pragmatic scheme
(called HFRED), that minimises the misfit between measured and model
data in a least-squares sense by varying the effective origin time of
penetration. Tests on numerically modelled data (synthetic
measurement) with known parameters showed that HFRED produced reliable
and accurate results.  However, the scheme has two major deficiencies:
\begin{enumerate}
\item The thermal diffusivity used for the sediment is computed from
  thermal conductivity according to a relationship proposed by
  \citet{Hyndman79}. This relationship has never been validated by
  experimental data and will certainly vary with sediment type.
\item The algorithm implemented in HFRED does not allow analysis of
  errors of the calculated undisturbed sediment temperatures and
  in-situ thermal conductivities in a rigorous way;  errors
  calculated by HFRED are always unrealistically low, compared to error
  estimates of about 5\% from other studies \citep{Hyndman79,Lister70}.
\end{enumerate}
To overcome these deficiencies and to incorporate platform independent
plotting routines, a mathematically sound inversion scheme of observed
temperature decays was implemented using Matlab$^{\circledR}$, a
widely used software package for numerical analysis. This allows
creation of very compact code for the inversion, on-screen graphics
and platform-independent plots.  In addition, automated processing or
reprocessing of a large number of individual measurements is possible.
The inversion of the integral describing the decay of a temperature
pulse (see equation~\ref{eq:thermal_decay_curve}) allows use of the
same algorithm for the calculation of undisturbed sediment
temperatures (using the frictional decay) and thermal conductivity of
the sediment (using the heat-pulse decay). Inversion theory allows
calculation of realistic errors in a well-defined and mathematically
rigorous way based on the sample rate and temperature resolution used.
Contrary to HFRED, the thermal conductivity and diffusivity are
treated as independent parameters, which may allow improvement of the
relationship of \citet{Hyndman79}.

In the following we propose an inversion scheme, coined
\ttc~(Temperature to conductivity), that was used to process the heat
flow measurements made on the German RV SONNE on the eastern flank of
Juan de Fuca during cruise SO111 in the summer of 1996
\citep{Villinger96}. In some concluding remarks we summarise our
experience up to now with the new reduction scheme.
%
%
\section{Theoretical background and inversion scheme}
\label{sec:theory}

The theoretical background for the analysis of heat flow measurements
described in the Introduction is discussed in \citet{Bullard54},
\citet{Lister79}, \citet{Hyndman79}, and \citet{Villinger87}. The
following simplified model for the sensor string is used: A cylinder
of radius $a$ and infinite extent in the $z$-direction is situated
in a homogenous infinite material. Whereas the material surrounding
the cylinder has a finite thermal conductivity $k$ and thermal
diffusivity $\kappa$, the cylinder itself is of infinite conductivity
and diffusivity, with the constraint that \rc, the product of specific
heat $c$ and density $\rho$ of the cylinder, remains finite. At time
$t=0$, the cylinder is at temperature $T_0$ and the ambient space at
$T_a$.  The temperature at the centre $r=0$ of the cylinder can then
be described by the thermal decay curve of the cylinder:
\begin{equation}
  \label{eq:thermal_decay_curve}
  T(\tau) = (T_0 -T_a) \cdot F(\alpha,\tau) + T_a
\end{equation}
with $F(\alpha,\tau)$ following \citet{Bullard54} and
\citet{Carslaw59}:
\begin{eqnarray}
  \label{eq:fat}
  F(\alpha,\tau) &=& \frac{4 \alpha}{\pi^2} \inlm{0}{\infty}{
  \frac{e^{-\tau u^2}}{u \phi(u,\alpha)} }{du} \\
  \phi(u,\alpha) &=& (u J_0(u) - \alpha J_1(u))^2 \nonumber \\
  && \label{eq:phi} + (u Y_0(u) - \alpha Y_1(u))^2
\end{eqnarray}
Here $\alpha = 2 (\rho c)_{s} / (\rho c)_{c}$ is the ratio of
the heat capacities per volume of the sediment and the cylinder, and
$\tau = \frac{\kappa}{a^2}t$ is the dimensionless time. $J_0$, $J_1$,
$Y_0$ and $Y_1$ are zero and first order Bessel and Neumann functions,
respectively and $u$ is the integration variable.
The temperature rise $T_0 - T_a$ at time $t=0$ can be expressed by
\begin{equation}
  \label{eq:Q_expressed_by_t-rise}
  T_0 - T_a = \frac{Q}{\pi a^2 (\rho c)_c}
\end{equation}
with $Q$ being the heat per unit length contained in the cylinder.
Equation \ref{eq:thermal_decay_curve} has an asymptotic solution that
approximates $F(\alpha,\tau)$ with 1\% accuracy for $\tau > 10$
\citep{Hyndman79}:
\begin{equation}
    \label{eq:long_time_sol}
    T(t) = \frac{Q}{4 \pi k t} + T_a 
\end{equation}
It is important to recall the limitations of this model by comparing
it with the sensor tube housing the temperature sensors:
\begin{enumerate}
\item The sensor tube is non-ideal, i.e.\ it has a finite
  conductivity and an internal structure.
\item The duration of the heating pulse is finite, usually in the
  order of 10--20s.
\item Axial heat flow will be inevitable but certainly small.
\item A thin water-layer between the sediment and the sensor tube
  may act as insulation to delay the achievement of thermal
  equilibrium.
\end{enumerate}
Measurements early in the temperature records will be inherently
affected by deviations from the model.  Therefore, these records have
to be excluded from analysis.  However, temperatures within the
analysed time range still show slight deviations from the ideal
behaviour.  This deviation can be best modelled by introducing a new
parameter, the time shift $t_s$. The measured time origin is always
the onset of the penetration or heat-pulse. Introduction of the
parameter $t_s$ approximates the heating of finite length by an
instantaneous temperature rise, shifted relative to the onset of the
heating. Although mathematically not rigorously proven, this concept
is reasonable from a physical point of view and has been shown to
provide reliable results \citep{Hyndman79,Villinger87}. The
justification for using $t_s$ will be investigated more thoroughly in
section~\ref{sec:synthetic_data} using a numerical model.

The goal of the processing scheme is to invert
equation~\ref{eq:thermal_decay_curve}. To achieve this, the actual
physical parameters have to be restored in equations~\ref{eq:fat}
and~\ref{eq:phi}. This yields
\begin{eqnarray}
  \label{eq:temp_decay_curve}
  T &=& \frac{8 k Q}{\pi^3 a^2 \kappa (\rho c)_c^2} \inlm{0}{\infty}{
    \frac{\displaystyle e^{-u^2 \frac{\kappa}{a^2}(t-t_s)}}{u
      \phi(u)}}{du}  + T_a \\
  \phi(u) &=& \left( u J_0(u) - \frac{2 k}{\kappa (\rho c)_c} J_1(u) \right)^2  \nonumber \\
   & & + \left( u Y_0(u) - \frac{2 k}{\kappa (\rho c)_c} Y_1(u) \right)^2
\end{eqnarray}
The equation assumes that the initial temperature $T_0$ at $t=0$ is
reached by introducing an amount of heat $Q$ either due to frictional
heating during penetration or a calibrated heat pulse.

According to equation (\ref{eq:temp_decay_curve}) the temperature
decay is a function of six parameters and time:
\begin{equation}
  \label{eq:vecT}
  T = T({\bf m},t) \quad \mbox{with} \quad {\bf m}=
  [k,\kappa,\Ta,Q,\rc,t_s]
\end{equation}
The inversion will be demonstrated for the general case of all six
parameters being unknown. In practical application of the technique,
the number of parameters needs to be reduced. This will be discussed
in section \ref{sec:processing_sequence}. 

The decay function is nonlinear
and cannot be inverted directly. One approach to solve this problem is
to expand $T({\bf m},t)$ in terms of a first order Taylor series
\citep{Menke89,Kristiansen82}.  The temperature will be expanded
around the ``true'' set of parameters ${\bf r}$, using an estimated
set of parameters ${\bf e}$:
\begin{equation}
  \label{eq:taylor}
  T({\bf r},t) = T({\bf e},t) + \sum\limits_{i=1}^{6} \left.
  \frac{\partial T}{\partial m_i} \right|_{m_i=e_i} \underbrace{(r_i
  - e_i)  }_{:=\Delta_i}
\end{equation}
This equation is linear in the difference vector ${\bf\Delta}$ and can
be inverted for this parameter vector. The left side of the equation
represents the data and the right side the model. Because the equation
is only of first order and hence not exact, a recursive scheme will be
used for the calculation of the true parameters whereby the result of
the $l$-th~iteration is used as an estimate ${\bf e}$ for the
$(l+1)$-th~iteration.
\begin{equation}
  \label{iterationscheme}
  {\bf e}^{(l+1)} ={\bf e}^{(l)} + {\bf\Delta}^{(l)}
\end{equation}
In general
\begin{math}
  {\bf e}^{(l+1)}
\end{math}
will be closer to the true parameter set than ${\bf e}^{(l)}$. Because
the parameters are of different orders of magnitude, the difference
vector ${\bf\Delta}$ has to be normalised so that only
relative weights appear in the model kernel \citep{Kristiansen82}.
\begin{equation}
  \label{eq:diffnormalize}
  \Delta'_{i} = \frac{\Delta_i}{e_i} \quad i=1,\ldots,6
\end{equation}
Applying this method to all data points for times $t_j\, (j=1,\ldots,N)$
yields a system of $N$~linear equations. These can be written in
matrix notation:        
\begin{eqnarray}\label{eq:short_matrix}
 {\bf G}{\bf \Delta}' &=&{\bf d} \\
  G_{j,i} &=& e_i \left. \frac{\partial T({\bf m},t_j)}{\partial
      m_i}\right|_{\bf m=e} \\
  d_j &=& T({\bf r},t_j) - T({\bf e},t_j)
\end{eqnarray}
${\bf G}$ is the model matrix or kernel, it contains no actual data,
only the assumptions of the model. ${\bf \Delta}$ is the model
parameter vector and ${\bf d}$ is the data vector.

The solution of the problem requires the inversion of the matrix
equation~\ref{eq:short_matrix} in order to obtain ${\bf \Delta}$.  We
use singular value decomposition (SVD) closely following
\citet{Menke89}. The decomposition of the model matrix produces three
matrices ${\bf G} = {\bf U} \cdot {\bf S} \cdot {\bf T^{T}}$ with the
solution to the inverse problem:
\begin{equation}
  \label{eq:svd_solution}
  {\bf \Delta_n} = {\bf V S^{-1} U^T d}
\end{equation}

\begin{figure}
  \begin{center}
    \includegraphics{\mygraphicspath 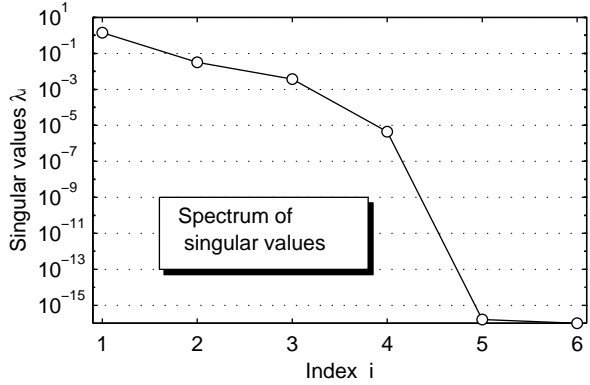}
    \caption[Spectrum of th SVD.]{Spectrum of the SVD for a
      representative set of parameters. For $i>4$, magnitudes are in the
      range of floating point accuracy of the implementation.}
    \label{fig:svd_spectrum}
  \end{center}
\end{figure}

Once conductivity-  and temperature-depth profiles are calculated
from the decay curves, heat flow values are computed using a method
proposed by \citet{Bullard39}.  Starting from the heat diffusion
equation for the horizontally layered case,
\begin{equation}
  \label{eq:heat_diff_pde}
  q = \frac{\partial}{\partial z} \left( k(z)T(z) \right)
\end{equation}
an integration over depth $z$ yields:
\begin{equation}
  \label{eq:thermal_depth_integral}
  T(z) - T_0 = q \underbrace{ \int\limits_0^z
  \frac{dz^{\prime}}{k(z^{\prime})}}_{ = w}
\end{equation}
The integral is called the thermal resistance $w$ and is calculated from
the conductivity-depth profile. A linear regression of the
temperature profile versus the thermal resistance gives the heat flow
density as the slope of
\begin{equation}
  \label{eq:bullard_linear_regression}
  T(z) = q\cdot w + T_0
\end{equation}
%
%
\section{Processing sequence}
\label{sec:processing_sequence}

In the last section a general inversion scheme was proposed which is
capable of inverting frictional and heat pulse decays.  It is
important that all parameters to be determined are independent of each
other. A measure of the linear dependence is the magnitude of the
diagonal values $\lambda_i$ of the singular matrix ${\bf S}$. A plot
of these values is called the spectrum of the singular value
decomposition and is shown for a typical temperature decay in
figure~\ref{fig:svd_spectrum}.  The accuracy of floating point figures
given by Matlab$^{\circledR}$ is about $10^{-16}$. The spectrum drops
to this level after the fourth singular value, suggesting that only
four independent parameters can be determined in the problem.
However, the accuracy and stability of the inversion is largely
influenced by the smallest included singular value and error bounds
are proportional to the inverse of the square of this value
\citep{Menke89}. Therefore, although it is possible to use four
independent parameters, in practice only three parameters were
determined to improve the results.  A thorough analysis revealed that
the singular value $\lambda_4$ is mostly influenced by $(\rho c)_c$.
This observation is a result of the heat capacity being a probe
parameter that is only poorly resolved by the model.  Therefore a
fixed value of 3.38~\uirc\ was used, based on the material and
geometry of the sensor string \citep{Nagihara93}. The possibility of a
systematic error resulting from a false assumption for the heat
capacity will be discussed later.
  
The restriction to three parameters for one inversion run leads to a
certain sequence of processing steps for a single penetration
\citep{Villinger87}:
\begin{enumerate}
\item Frictional decay: $Q,T_a,t_s$ are calculated. $k,\kappa, (\rho c)_c$
  are held fixed, using estimated values for $k$ and $\kappa$.
\item Heat pulse decay: The residual of the frictional decay is
  subtracted from the heat pulse decay thus reducing the ambient
  temperature $T_a$ to a known value of zero. The inversion is used to
  compute $k,\kappa,t_s$ with $T_a, (\rho c)_c, Q$ held constant. $Q$
  is known as an input parameter and $T_a$ is zero after the
  reduction.
\item The calculations in steps (i) and (ii) can be repeated, this time
  using the calculated values of $k$ and $\kappa$ as
  improved estimates in step (i). 
\end{enumerate}
\begin{table}
  \begin{center}
    \begin{tabular}[t]{c|c|c|c|c|c|}
      && \multicolumn{2}{|c|}{Frictional Decay} &
      \multicolumn{2}{|c|}{Heat-pulse Decay} \\\hline
      && Value & Error & Value & Error \\\hline\hline
      $k$ & $\ucb$ & 1.0 &- &1.0 & 0.018 \\\hline
      $\kappa$ &  $\udb$ & 2.0&-&2.0&0.22 \\\hline
      $Q$ & $\uqlb$& 30 & 9.6 & 600 & - \\\hline
      $(\rho c)_c$ & $\urcb$ &3.38 & - & 3.38 & - \\\hline
      $t_s$ &[s] & 1 & 37 & 50 & 1.9 \\\hline
      $T_a$ &[mK] & 500 & 1.4 & 0.0 & - \\\hline
    \end{tabular}
    \caption{Values of the parameters used to calculate a model kernel
      for frictional and heat pulse decay. Errors are calculated from
      equation \ref{eq:model_error}. For the fixed parameters in the inversion
      no error is given.} 
    \label{tab:used_parameters}
  \end{center}
\end{table}
The technique was first used on the results of research cruise SO111
(see section~\ref{sec:results_so111}). Penetrations took place over
young crust and comparatively warm sediments. Especially for the upper
thermistors on the sensor string it was noticed that the frictional
heat of the penetration did not suffice to raise the temperature from
seawater temperature above the ambient sediment temperature. This
caused very small signals for the frictional decay of some of the
sensors, for example T11 in figure~\ref{fig:penexample}.

For the model we used it turns out that such a small frictional decay
can be described equally well by a large heating shifted by a large
amount in time or a small heating and a small time shift. This
ambiguity in the parameters $Q$ and $t_s$ cannot be resolved without
additional information and leads to problems when inverting frictional
decays.  Because the important parameter $T_a$ is not influenced by
this ambiguity, a practical approach to this problem is to set the
time shift to a fixed value, namely zero for these cases.

To determine occurences of this problem, the first order approximation
(equation~\ref{eq:long_time_sol}) is used to determine decay curves
with very little frictional heat. If $Q$, computed from this equation,
is below a certain threshold, $t_s$ is set to zero.

The stability of the inversion of heat pulse decays is affected by
$\kappa$-values that fluctuate from iteration to iteration. In the
worst case this causes the algorithm to diverge. To prevent this, only
small changes in $\kappa$ are allowed between iterations, effectively
damping any oscillations.  This constraint is used only for the first
three iterations.

\begin{figure}
  \begin{center}
    \includegraphics{\mygraphicspath 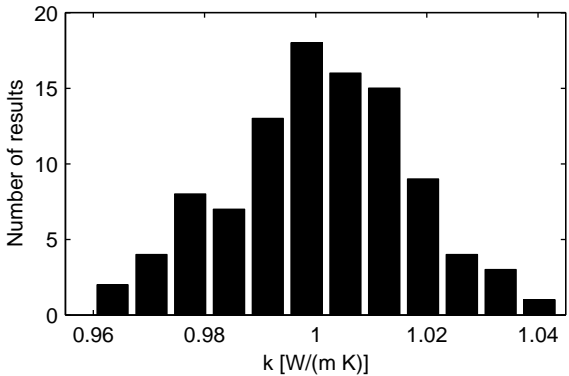}
    \caption[Distribution of conductivity for noisy data]{Distribution
      of conductivity values when synthetic, normally distributed noise
      with $\sigma$ = 1~mK is added to a synthetic data set. The
      mean value is $(1.001\pm0.017)$\uic.}
    \label{fig:conductivity_distribution_noisy_data}
  \end{center}
\end{figure}

%
%
\section{Calculation of errors}
\label{sec:error_calculation}

For inversions theoretical errors can be calculated using the
model kernel ${\bf G}$ \citep{Menke89}:
\begin{equation}
  \label{eq:model_error}
  \bf cov_m = G^{-I} cov_d G^{-IT}
\end{equation}
Here ${\bf cov_m}$ and ${\bf cov_d}$ are the covariance matrices of
the model parameters and the data, respectively.  ${\bf G^{-I}}$ is
the inverse of the kernel, in our case ${\bf VS^{-1}U^{T}}$
(equation~\ref{eq:svd_solution}). This equation is only exact for
linear problems with normally distributed data.  Nevertheless, it can
be applied to a nonlinear problem, if the problem can be approximated
by a linear function in the vicinity of the solution. It is a useful
property of equation~\ref{eq:model_error} that no data is used for the
calculation of errors, only the data covariance and a model kernel are
needed. The kernel can be computed using
equation~\ref{eq:short_matrix} by assuming a parameter vector with
representative values. This choice is not critical since the kernel
and corresponding errors vary only slowly with the parameters.  This
feature of equation~\ref{eq:svd_solution} is particularly useful for
design studies and theoretical evaluation of the limitation of a
certain model.

The covariance matrix of the data is not known a priori but in our
case it can be assumed that data errors are uncorrelated and mainly
due to the finite temperature resolution of $\pm$1~mK. ${\bf cov_d}$
is then a matrix with temperature variances on its main diagonal and
zeros elsewhere. To compute the kernel ${\bf G}$ using
equation~\ref{eq:short_matrix}, a modelled decay curve from 120--420~s
with 10~s sampling rate and two sets of parameters for frictional and
heat pulse decay are used, respectively. These values, together with
the calculated errors, are summarised in
table~\ref{tab:used_parameters}.  For the frictional decay, errors for
$Q,t_s$ and $T_a$ are determined, according to the processing sequence
given in section~\ref{sec:processing_sequence}.  The ambient
temperature can be resolved with a standard deviation of 1.4~mK, or
with a relative error of 0.28\%. For the heat pulse decay, errors for
$k,\kappa$ and $t_s$ were computed. The standard deviation of the
thermal conductivity is 0.018~\uic\ (1.8\%).

The error bounds given in table~\ref{tab:used_parameters} give a
general idea of the errors to be expected when using the algorithm.
The relevant parameters conductivity and sediment temperature can be
calculated with relative errors of about 2.0 and 0.5\%, respectively.
For the parameter range encountered in practice, errors will vary
slightly. Therefore, for each penetration errors are calculated from
the inversion results and stored together with them.
%
%
\section{Tests of the proposed inversion scheme}
\label{sec:synthetic_data}

To verify the theoretically derived error bounds, normally distributed
noise with $\sigma$ = 1 mK was added to the exact model data. This
data was then inverted. To obtain reasonable statistics the procedure
was repeated 100~times, producing results for $k$ which are shown in
the histogram in
Figure~\ref{fig:conductivity_distribution_noisy_data}. The mean of the
computed conductivities, (1.001$\pm$0.017)\uic, agrees very well with
the expected value of (1.00$\pm$0.018)\uic.

The value of \rc\ is only known on the basis of geometric
considerations \citep{Nagihara93}. Additionally, its value could vary
by a few percent over the length of the sensor string because of more
wiring in the upper parts of the string. If a wrong value for \rc\ is
assumed it will influence the results of the inversion in a systematic
way. Therefore it is useful to investigate the magnitude of the error
introduced by an incorrect value.

\begin{figure}
  \begin{center}
    \includegraphics[width=8cm]{\mygraphicspath 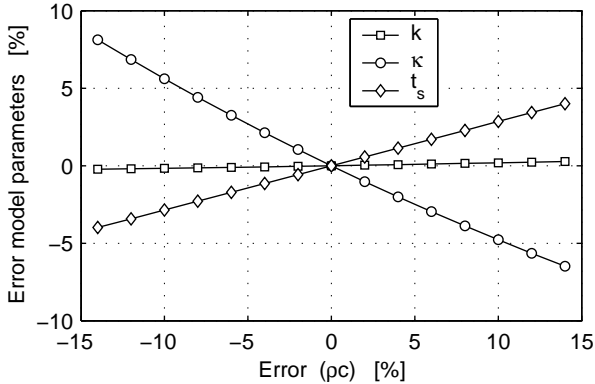}
    \caption[Influence of false \rc-values on inversion]{Influence of
      false \rc-values on inversion results. Variations of up to 15\%
      result in no significant change in $k$.}
    \label{fig:wrong_rc}
  \end{center}
\end{figure}
A synthetic temperature record was calculated using the same
parameters as in the previous section. \rc\ was chosen to be
3.38~\uirc, the value given by \citet{Nagihara93}.  This record was
the input to an inversion in the heat pulse-configuration ($k$,
$\kappa$, $t_s$\ are calculated). Several runs were made, each time
with a slightly different value for the heat-capacity, ranging from
2.9 to 3.9\uirc which corresponds to a relative deviation of about
$\pm$15\%.

Figure~\ref{fig:wrong_rc} shows relative errors in the calculated
model parameters when \rc\ is varied. For the maximum bias, the
deviation of the conductivity is only about 0.2\%. This value is
considerably smaller than the errors introduced from other sources.

The situation is different for the other two parameters though. The
systematic deviations are significant compared to the random errors.
$t_s$\ only has  meaning within the framework of the inversion
algorithm, but $\kappa$ is a physical parameter and a relationship
based on the calculated $\kappa$ could be in error by a few percent.

The tests were conducted using synthetic data based on
equation~\ref{eq:temp_decay_curve}. A time shift can be computed for
model curves but it cannot be verified whether $t_s$\ is able to
approximate the non-ideal parts of real temperature records.

A closer approximation to reality is a numeric model that is able to
model finite heat pulse length and finite probe conductivity.  For
this reason a program called TFELD was used which is based on an
algorithm proposed by \citet{Villinger85}. This program allows
modelling of heat diffusion in a cylindrically layered space with
various heating functions. Each of the cylindrical layers is
characterised by a set of physical parameters $\rho$, $k$ and $c$. In
this case a model with only two layers was employed, the inner and the
outer layer representing the probe and the sediment, respectively.

In the first model used the conductivity and diffusivity of the probe
(layer 1) were set to 1000\uic\ and 2000\uid\ respectively to give a
good approximation of an ideal probe. The ambient sediment (layer 2)
values of 1\uic\ and 2\uid\ were used as representative values for
conductivity and diffusivity. An instantaneous temperature rise of the
inner cylinder was used as the heat source. The computed temperature
decays were used as input for the inversion algorithm. The results are
shown in table~\ref{tab:tfeld_results}.
\begin{table}
  \begin{center}
    \begin{tabular}{|lc|c|c|} \hline
      && TFELD& T$_2$C \\\hline\hline
      $k$     &\ucb  & 1.0 & 1.0002 \\\hline
      $\kappa$&\udb  & 2.0 & 2.0019 \\\hline
      $t_s$     &[s] & 0.0 & 0.0016 \\\hline
    \end{tabular}
    \caption[Inversion results for the TFELD model.]{Inversion results
      of the TFELD model. The results show good agreement.}
    \label{tab:tfeld_results}
  \end{center}
\end{table}
As expected the inversion gives the correct result within the
precision of the numerical computations.

\begin{figure}
  \begin{center}
    \includegraphics{\mygraphicspath 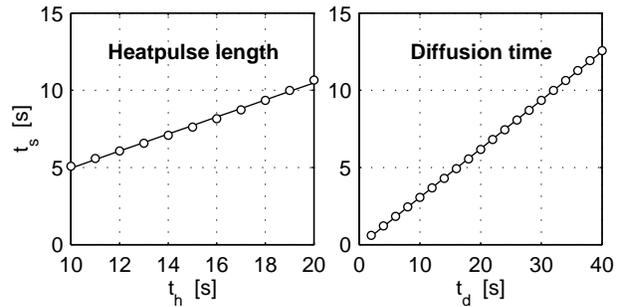}
    \caption[Influence of non ideal probe parameters.]{Influence of
      non ideal probe parameters: a) finite heating b) finite
      diffusivity.}
    \label{fig:system_dev}
  \end{center}
\end{figure}

The main objective of using the TFELD model was to study the influence
of a non-ideal probe on the temperature decay curves. In the
literature \citep{Hyndman79,Villinger87,Nagihara93} it is assumed that
the effects of the finite thermal conductivity of the cylinder and the
finite heat pulse on the temperature decay will be combined in the
time shift $t_s$.

In order to separate effects of finite heating and probe geometry two
experiments were conducted, one with an approximately ideal probe and
finite heat pulse and a second with real probe geometry and
infinitely short heat pulse.

For the first experiment probe conductivity was fixed at 1000\uic{}
and heat pulses with duration from 10--20~s were used as input
to TFELD. The resulting decay curves were inverted using \ttc{}. If
the time shift from the inversion is plotted versus the length of the
heat pulse (Figure~\ref{fig:system_dev}a), a strong linear
relationship can be seen.

A similar test was run to determine the relationship between probe
geometry and time shift. It was found that the parameter suited best
to describe the probe geometry is the diffusion constant
$t_d$:
\begin{equation}
  \label{eq:diffusion_constant}
  t_d = \frac{a^2}{\kappa}
\end{equation}
This value describes the time that a temperature disturbance would
take to travel the distance $a$. The diffusion time of the probe was
varied from 2--40~s in this experiment corresponding to values for $k$
and $\kappa$ of 40--2~\uic\ and 80--4~\uid, respectively. The initial
temperature field was $T_0$ inside the probe and zero outside.  Again
these model curves were inverted and the resulting time shifts plotted
versus the diffusion time to obtain the linear relationship seen in
Figure~\ref{fig:system_dev}b.

If a linear fit is calculated for both curves, the following
relationships are obtained:
\begin{eqnarray}
  t_s = 0.55 t_h - 0.56\,\qquad\mb{[s]}   \label{eq:timeshift_fit_hp} \\
  t_s = 0.31 t_d - 0.08\, \qquad\mb{[s]} \label{eq:timeshift_fit_td}
\end{eqnarray}
Here $t_h$ is the duration of the heat pulse. It is intuitively
explicable that a temperature record, generated by a finite heat pulse
can be best described by a perfect pulse shifted half the length of
the original pulse.  Similarly the probe geometry will result in a
time shift that is directly dependent on the time the heat pulse needs
to travel through the internal structure. This is reflected in
equations~\ref{eq:timeshift_fit_hp} and \ref{eq:timeshift_fit_td},
respectively.

One reason for the development of our algorithm was to evaluate the
possibilities of calculating the conductivity directly from the
frictional decay. In this case the only known parameter is $(\rho c)_c$.
From the discussion of the spectrum of the SVD, it is clear that a
reduction of the number of parameters to at least four is necessary to
invert equation~\ref{eq:thermal_decay_curve} as an overdetermined
problem. To reduce the number of parameters in the problem by one, a
relationship between conductivity and diffusivity can be
used \citep{Hyndman79}.
\begin{equation}
  \label{eq:kappa_Hyndman}
  \kappa = \frac{k}{5.79-3.67k+1.016k^2}\qquad
  \left[ 10^{-6} \frac{\mbox{m}^2}{\mbox{s}} \right]
\end{equation}
This assumption is justified because both conductivity and diffusivity
of marine sediments are mainly porosity controlled. Using
equation~\ref{eq:model_error} one can calculate expected error bounds
for this configuration. If an observational error of $10^{-3}$~K and
reasonable parameters for probe and sediment are assumed, the errors
in the model parameters are as follows:
\begin{itemize}
  \item[] $\Delta k$ = 6.6\uic
  \item[] $\Delta Q$ = 1500\uiql{}
  \item[] $\Delta t_s$ = 58~s
  \item[] $\Delta\Ta$ = 12~mK
\end{itemize}
It can be seen in equation~\ref{eq:model_error} that the model
covariance is linearly related to the data covariance. If an error
margin of $0.05 \uic$ ($\approx \pm5$~\% for marine sediments) is
assumed as an acceptable error level for the conductivity, this means
a reduction of $\Delta k$ by 2 orders of magnitude. This means as well
a reduction of the observational error by 2 orders of magnitude. Thus,
accuracy of temperature readings has to be reduced to an extremely low
and unrealistic level of $\pm10^{-5}$~K.  The results of these tests
confirm that the frictional decay is not sensitive to the thermal
conductivity of the surrounding sediment and a successful inversion
for $k$ with realistic error bounds for the temperature measurements
is not possible.

The reason for this failure is implicit in the model. Not enough
information is provided by a frictional decay curve to resolve the
ambiguity of the problem using an inversion scheme. Programs exist,
however, that are capable of modelling the full frictional decay
(forward modelling), including the early times
\citep{Lee94,Villinger85}. This allows other constraints to be added
to the problem, thereby reducing the ambiguity \citep{Lee94}.  It has
to be considered, however, that any additional assumptions and
a-priori information will directly affect computation of thermal
conductivities and have to be chosen carefully.

\begin{table}
  \begin{center}
    \begin{tabular}[t]{|l|l|} \hline
      \multicolumn{2}{|l|}{\bfseries Thermal parameters:}\\\hline
      Conductivity & (1.609 $\pm$ 0.016)\uic \\\hline
      Diffusivity & 7.3\uid\\\hline
      Heat capacity & 2.20\uirc\\\hline\hline
      \multicolumn{2}{|l|}{\bfseries Measurement parameters:}\\\hline
      Heating power & 8, 12, 15\uipl\\\hline
      Heat pulse duration & 5, 10 s\\\hline
      Sampling rate      & 0.5 s \\ \hline
      Observational error & 10$^{-4}$ K
    \end{tabular}
    \caption[Parameters for needle probe measurements.]{Upper half:
      Thermal parameters of the material used for needle-probe
      measurements. Lower half: Parameters of the different records
      taken.}
    \label{tab:params_needle_probe_measurements}
  \end{center}
\end{table}

%
%
\section{Inversion of needle-probe measurements}
\label{sec:needle_probe_measurements}

To test the algorithm with real data, a set of pulsed needle-probe
measurements were used.  These were made in a ceramic alloy with a
known thermal conductivity close to the value of deep sea sediments.
The data were kindly provided by TeKa Inc.~(Berlin, Germany), a
company specialised in thermal conductivity measurements and
equipment. The true conductivity of the ceramics was measured with a
divided bar device (TeKa Inc., pers.\ comm.).  The relevant parameters
of the material and the recording settings are summarised in
table~\ref{tab:params_needle_probe_measurements}.  For each
combination of parameters three decay curves were recorded, resulting
in a total of 18 decay curves.

\begin{figure}
  \begin{center}
    \includegraphics{\mygraphicspath 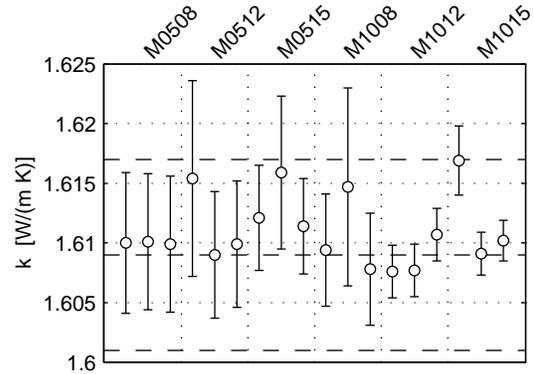}
    \caption[Inversion results needle-probe records]{Inversion results
      for the needle-probe measurements.  The circles represent
      calculated conductivities with 1-$\sigma$ error bars for the
      18~records.}
    \label{fig:conductivity_results_needle_probe}
  \end{center}
\end{figure}

A drift correction was applied to the temperature records to reduce
the value of the ambient temperature \Ta\ to zero. The heat capacity
of the probe was unknown so that the decay curves were inverted using
four unknown parameters: $[k,\kappa,\rc,t_s]$. The results for the
conductivity with 1-$\sigma$ error bars are shown in
figure~\ref{fig:conductivity_results_needle_probe}. The circles
represent the computed thermal conductivities together with 1-$\sigma$
error bars. The dashed lines are the mean thermal conductivity and 1\%
error margins determined with the divided bar method. Shown are all 18
measurements in six groups of three measurements with the same heating
parameters. On the top horizontal axis, the heating parameters are
encoded in the names: The first two digits represent the duration [s],
the last two digits the heating power per unit length [W/m] of the
heat pulse. For each set of parameters, three measurements were made.

There is excellent agreement between expected and calculated values of
the heat conductivity. Errors of $k$ were between 0.2 and 0.5\%. For
the $\kappa$-values the errors are 3--15\%\ and for $(\rho c)_c$
10--30\%.  The high accuracy of the conductivity value compared to
previous tests is due to the small observational error of the needle
probe. As stated before, large errors for $\kappa$ and $(\rho c)_c$
are related to the inversion with four unknown parameters.

The error bounds generally decrease to the right in
figure~\ref{fig:conductivity_results_needle_probe}. Changes in length
of the heat pulse and heating power affect the total heat input and
the recorded temperature signal. Thus it appears that calculated error
bounds vary with the magnitude of the recorded temperature changes, as
expected.

\begin{figure}
  \begin{center}
    \includegraphics{\mygraphicspath 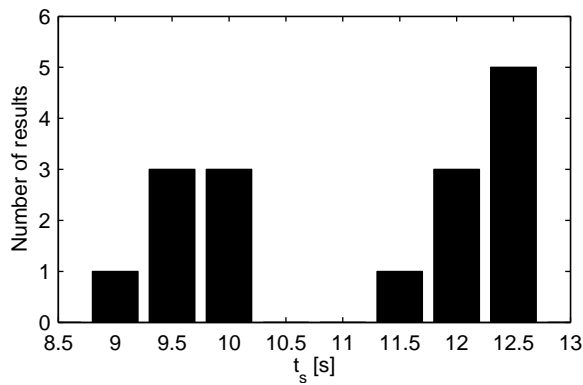}
    \caption[Distribution of time shifts for the needle-probe
    records.]{Distribution of calculated time shifts for the
      needle-probe measurements. Two clearly distinguishable groups
      can be identified that are about 2.5~s apart from each other.}
    \label{fig:timeshift_needle_probe}
  \end{center}
\end{figure}

It is interesting to check whether the hypothesis that the time shift
is connected to the diffusion time via a simple relationship of the
type in equation~\ref{eq:timeshift_fit_hp} holds true for real data. 
Figure~\ref{fig:timeshift_needle_probe} shows that all $t_s$ fall into
two clearly distinguishable groups according to the duration of the
heat pulse (5 and 10~s). The difference of about 2.5--3.0~s is in
good agreement with the expected value, which should be half the
difference of the lengths of the heat pulses.
%
%
\section{Results from research cruise SO111}
\label{sec:results_so111}

During summer of 1996 cruise SO111 was conducted off Western Canada
on the German research vessel SONNE.  The objective was to study the
effect of hydrothermal circulation on marine heat flow on the Eastern
flank of the Juan de Fuca Ridge. The survey area was located near the
Cobb Offset at about 47$^\circ$ 30' N, 129$^\circ$ 0' W.  During the
cruise 8~stations with 104~successful penetrations were made.  For a
detailed discussion of the measurements during this cruise see
\citet{Villinger96}.

This cruise provided the first instance to use the program on a
regular basis.  The following is concerned mainly with the application
of the inversion algorithm. Though all penetrations were inverted
successfully using the described algorithm only data from station~2
with 30~penetrations will be used to illustrate the applicability of
the program.

In the first stage of processing the moment of penetration of the
probe and the onset of the heat pulse have to be picked manually using
the raw temperature data. The data can then be input to the inversion
program that calculates thermal parameters according to the previously
described sequence. The thermal parameters are then in turn used to
calculate heat flow values.

To minimise the influence of the non-ideal probe at early times, only
temperatures for times $>$ 120~s relative to the picked origin time
were used. The value for this start time depends on the probe geometry
and was found empirically by examining the fit for several start
times.  For example \citet{Nagihara93} used a sensor string with a
diameter of 9.52~mm versus 8.0~mm in our design and a value of 200~s
as the starting time for their analysis.

\begin{figure}
  \begin{center}
    \includegraphics{\mygraphicspath 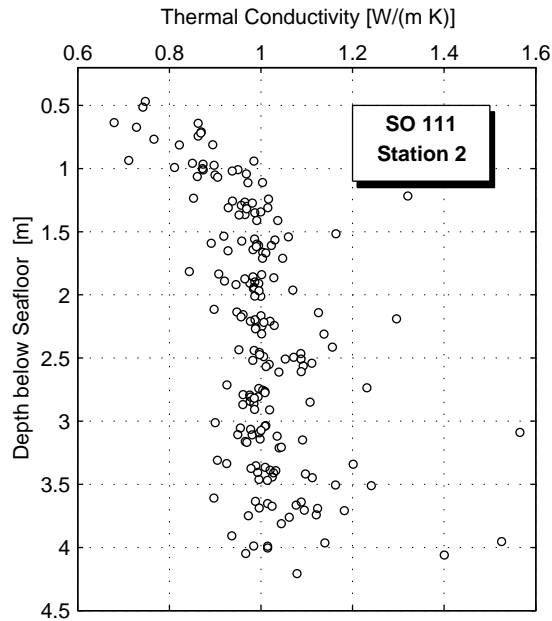}
    \caption{Conductivity-depth profile for station~2.}
    \label{fig:conductivity_profile}
  \end{center}
\end{figure}

After calculation of conductivities and thermal gradient, the absolute
penetration depth can be calculated using the bottom water temperature
and the assumption that the temperature is continuous at the
sediment-water interface.  Figure~\ref{fig:conductivity_profile} shows
the conductivity versus the sensor depth for station~2.  A strong
increase in conductivity over the first metre can be seen that is
possibly caused by a decrease of porosity in the surface sediments.
The outliers in the lower part of the plot might be attributed to a
partially penetrated sand layer of varying depth. This corresponds to
the observation that often the penetration was stopped by a layer of
harder material.

\begin{figure}
  \begin{center}
    \includegraphics{\mygraphicspath 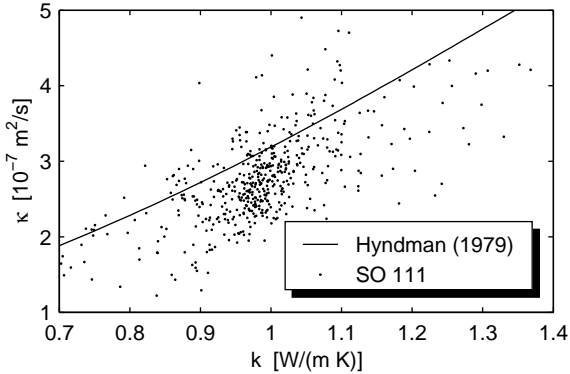}
    \caption[Calculated diffusivity versus conductivity for
    station~2.]{Calculated diffusivity versus conductivity for
      station~2. The values are compared to an empirical relationship
      by \protect\citet{Hyndman79}.}
    \label{fig:hyndman}
  \end{center}
\end{figure}

Previous workers (\cite{Bullard54, Ratcliffe1960, VonHerzen59,
  Hyndman79}) determined thermal diffusivity by measuring
conductivity, density and porosity.  Assuming that heat capacity and
conductivity are controlled mainly by the water content, they
constructed a relationship between thermal conductivity and
diffusivity.  In our approach, thermal diffusivity is computed as an
independent parameter in the inversion.  Figure~\ref{fig:hyndman}
shows calculated diffusivity values compared to the relationship
(Equation~\ref{eq:kappa_Hyndman}) given by \citet{Hyndman79}. The
inversion results are significantly lower than expected by the
theoretical curve. To our knowledge no other measurements of our type
for marine sediments are published to verify the results.  It is
apparent, though, that a simple equation is not sufficient to describe
the variations in the relation between $k$ and $\kappa$.

\begin{figure}
  \begin{center}
    \includegraphics{\mygraphicspath 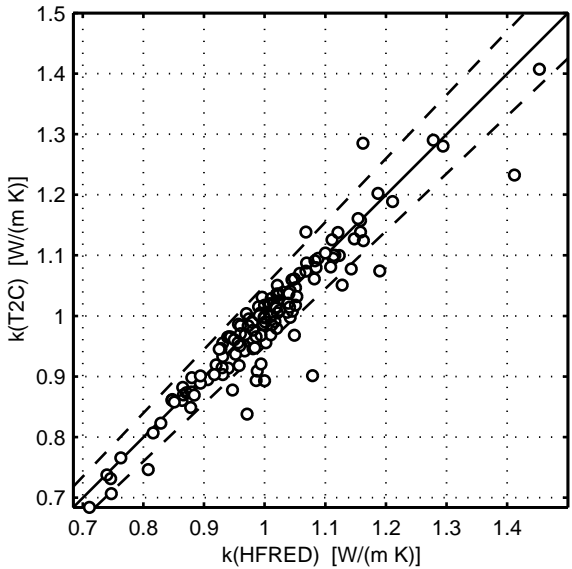}
    \caption[Comparison between T$_2$C and
    HFRED]{Comparison of thermal conductivities between \ttc{}
      described in this paper and HFRED \citep{Villinger87}. Dashed
      lines show a relative deviation of $\pm$ 5\%.}
    \label{fig:compare_hfred}
  \end{center}
\end{figure}

It is instructive to compare the results of HFRED \citep{Villinger87}
to our results.  Figure~\ref{fig:compare_hfred} shows differences in
computed conductivity for both algorithms. A high agreement is visible
with deviations usually within $\pm$5\%. 

\begin{figure}
  \begin{center}
    \includegraphics{\mygraphicspath 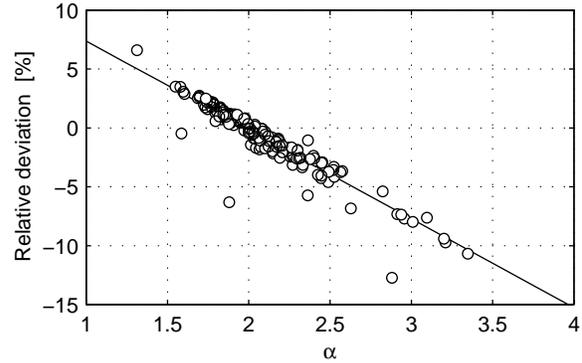}
    \caption[Conductivity differences as a function of
    $\alpha$]{Relative differences of conductivities computed with
      HFRED and \ttc{} as a function of $\alpha$.}
    \label{fig:compare_hfred_alpha}
  \end{center}
\end{figure}

HFRED uses a constant value for the ratio of the heat capacities
$\alpha$ of 2.0 because of the lack of knowledge of the thermal
diffusivity. In the \ttc{} algorithm $\alpha$ is not used but can be
calculated from the computed thermal diffusivity as $\kappa = k /
\rc$.  Figure~\ref{fig:compare_hfred_alpha} shows relative differences
of the thermal conductivity versus $\alpha$. It is apparent that large
deviations from $\alpha=2$ correspond to large differences in thermal
conductivity, suggesting that a constant value of $\alpha$ is not
sufficient to describe all measurements accurately.

\section{Conclusion}
\label{sec:conclusion}

An algorithm is described to invert heat flow measurements made with a
violin-type heat probe including in-situ thermal conductivity
measurements after the pulse method of \citet{Lister70}. A theoretical
analysis of the inversion algorithm shows that undisturbed sediment
temperatures can be determined with an error of about 1--2~mK. The
absolute error of the in situ thermal conductivity $k$ is less than
0.002~\uic. It is also possible to compute thermal diffusivities, but
with considerably higher errors of about 10~\%.

Numerical experiments with synthetic temperature decay data reveal a
strong relationship between the time shift and the thermal probe
parameters which can be explained in a quantitative way by the finite
length of the heat pulse and the diffusion time constant of the sensor
string.

Measured data, obtained with a needle probe in material with known
thermal conductivity, confirm the accuracy of the inversion procedure
and show that the algorithm is suited to the analysis of pulsed
needle probe measurements. Our results show that is is possible to
succesfully use the algorithm in pulsed needle probe measurements.

The inversion of the data obtained on SO111 proves that the described
algorithm is robust and well suited for automated processing of a
large number of heat flow penetrations. The embedding of the software
in a suite of mathematical software allows simple further analysis of
the data and easy development of additional tools. The relative
accuracy of our thermal conductivity results is in the range of
1--3\%. Undisturbed sediment temperatures can be computed with
relative errors of 0.5--1\%.  The comparison with results obtained
with the previously used program HFRED \citep{Villinger87} shows good
agreement between both algorithms. Deviations are generally due to the
assumption of $\alpha=2$ by HFRED.

\bibliography{gz358}
\bibliographystyle{gji} 

\begin{acknowledgments}
  
  The paper was improved by the thorough reviews of our colleagues,
  Ingo Grevemeyer, Norbert Kaul, and Marion Pfender as well as R. von
  Herzen and an anonymous reviewer. The research cruise SO111 was
  kindly funded by the German Ministry for Research and Technology,
  Grant No.~03G0111A.

\end{acknowledgments}

\end{document}